# Quantum Anti-Zeno Treatment of Zeno-type Sleep Disorders

Rajat Kumar Pradhan*
Rajendra College, Bolangir, Odisha, India-767002

(Date: 19.11.2011)

## Abstract

It is proposed that for those sleep disorders of psychological origin which can be considered to be a Quantum Zeno Effect-type phenomenon of *persistence in the waking state* due to the inhibition of the transition to the Deep Sleep state, the treatment may very well lie in the application of the principle of *accelerating the decay by the introduction of a third state* which facilitates the transition as realized in the Quantum Anti-Zeno effect. Steps of practical therapeutic implementation of the program are delineated.

*email: rajat@iopb.res.in

## 1. Introduction

Quantum theory[1,2,3] is gradually proving to be the most versatile and the most successful framework for the investigation of diverse phenomena in the realms of matter as well as mind. It has been proposed that the well-known Quantum Zeno Effect (QZE)[4,5] plays the most important role in establishing the mind-brain relationship[6,7] through continued attention. That the phenomena in the planes of the physical and the psychical have a lot of parallels and that they can be described by very similar methods was realized long ago by Pauli and Jung[8]. It has also been proposed recently by the present author[9] that the three states of consciousness, viz. waking, dreaming and sleep, experienced daily may be described by the composition of two spin-like quantum mechanical observables characterizing the subject and the object. The formulation is such that it allows for finer experiential modes or levels within each state or band. For example, one can say that the experiential state $|\omega(\theta_i,t)\rangle$ at time t is that of the $i^{th}$ thought-form $\theta_i$. In the waking state $|\omega_1(\theta_i,t)\rangle$ each thought-form has an objective counterpart $\Theta_i$ in the external physical world whose neural correlate is perceived as that thought-form or mental image $\theta_i$. Similarly, we can also characterize the dream state at time t as $|\omega_2(\theta_i',t)\rangle$ ( 't' is the waking time) when the dream-form $\theta_i'$ corresponding to $\Theta_i$ is experienced. The continuous streams of such thought-forms make up our daily experience in the waking and dream states.

However, the state of deep sleep $|\omega_3\rangle$ happens to be different since it is characterized by non-perception of either the internal thought-forms or of their external objective counterparts. It is an objectless and thoughtless state of ignorance or unconsciousness that we call deep sleep. The period spent by a healthy adult in deep sleep is roughly 6-7 hrs/day i.e. about one-third of the period spent in non-sleep. The transition to sleep and also the emergence therefrom is usually through the intermediate state of dream $|\omega_2\rangle$ in healthy and normal persons although the direct route is also available[9]. Insomnia (sleeplessness) then becomes a phenomenon in which the transition from $|\omega_1\rangle$ to $|\omega_3\rangle$ through the direct and the dream routes is inhibited due to any one or more of the three factors: (a) Environmental factors (b) Physiological factors and (c) Psychological factors.

In this article, we shall assume that the first two factors are either absent or have been fully taken care of and that the sleep disorder is due only to the psychological causes i.e. the patient tries to sleep at the appropriate time but because of some thoughts running riot in his head he is unable to 'switch himself off' to move into dream and consequently to sleep. (It is to be remembered that normally the pre-sleep dreams are not registered as experience unless some disturbance wakes one up in the threshold of one's entry into deep sleep.). Just as a person can come directly to $|\omega_1\rangle$ from $|\omega_3\rangle$ when there is some sudden disturbance (external noise or some other such forceful sensory input), so also the sleeping pills can enforce the direct transition to $|\omega_3\rangle$ from $|\omega_1\rangle$ which is more important in the treatment in acute cases. This is because the dream route requires $|\omega_4\rangle$, the fourth state, which is even more difficult to pass on to compared to $|\omega_3\rangle$ in the case of patients with acute sleep disorder. However, once the acuteness is reduced by medication, the case is not cured but reduces to chronic insomnia.

Thus, we assume that once the patient is led into $|\omega_2\rangle$, the transition to $|\omega_3\rangle$ is automatic and smooth and the central problem then becomes one of effecting the transition from $|\omega_1\rangle$ to $|\omega_4\rangle$ which will lead naturally to $|\omega_2\rangle$ and from then on to $|\omega_3\rangle$.

We propose that this particular type of insomnia can be modeled as a QZE-type inhibition of the transition from $|\omega_1\rangle$ to $|\omega_4\rangle$ due to the persistent experience of the rioting thoughts in the waking state. This persistent experience of the waking state is the counterpart of inhibition of transition due to continuous measurement in QZE. We propose further a non-medicinal treatment of the same disorder basing on the inverse phenomenon of Quantum Anti-Zeno Effect (QAZE) [10,11], wherein the intentional introduction of an auxiliary state $|\omega_1(\theta_a, t)\rangle$ facilitates this transition.

In section-2 we paraphrase the problem of stress-induced insomnia as a QZE and in section-3 we present the proposed solution via QAZE. In section-4 practical steps of eliminating the environmental and physical factors and of aiding the QAZE transition are delineated which can be practiced under expert supervision for the first one week or so individually or in groups. Afterwards, the patients may be asked to continue the method of perfecting the QAZE-transition themselves. In section-5 we conclude with a discussion of the limitations of the method and other possible applications. The whole article is written in a non-technical manner so that it is easily accessible to doctors, psychiatrists as well as to the patients and laymen.

## 2. Insomnia as Quantum Zeno Effect

The majority of secondary insomnia cases happen to be due to psychological factors which delay the onset of sleep beyond limit and the patients complain of uncontrollable thoughts running riot the moment they retire to bed and close their eyes. The problem gets further complicated by the apprehension of recurrence of the same phenomenon every night and this weakens the will of the patients considerably and they complain of recurrent sleepless nights i.e. chronic insomnia. In most of the cases, it is found that the kinds of thoughts that bother them are mostly those of some past or future trouble. It is only very rarely that one gets sleep disorders due to too much of pleasant experiences. Pleasant experiences, fancies and fantasies usually usher in good sleep! It is the unpleasant experiences like disasters and debacles, fears and phobias, failures and frustrations, insults and abuses, mishaps and misfortunes, losses and bereavements, traumas and tortures etc. which one has undergone or is likely to undergo that come to haunt the patient on the bed. The patient may struggle for hours together without getting any semblance of sleep and this phenomenon in chronic cases may continue even beyond a month or so.

This is the case that can be taken up as a classic illustration of Quantum Zeno Effect in the psychological domain, where the continuous cognizance of the problematic thoughts $\theta_p$ by the consciousness prevents the momentary slide into the fourth state which could have brought the dream state on, and which in its sequel would have ushered in deep sleep. It is proposed here that this inhibited transition from $|\omega_1(\theta_p, t)\rangle$ to $|\omega_4\rangle$ due to continuous observation can be removed by recourse to the inverse process of Quantum Anti-Zeno Effect.

In fact, it seems that all cases of insomnia, including the primary ones of unknown causes, whether chronic or acute, can be considered as being effectively due to this kind of QZE and correspondingly can be cured using QAZE. It is clear that the proposed solution will have to be of the psychological type, and hence, non-medicinal, but one that can easily be practiced by anyone suffering from any kind of insomnia. In this respect, when it comes to the choice of an auxiliary state in QAZE for fixing the attention on, it is worth emphasizing that of the many involuntary processes that continue throughout the day including the period of deep sleep the most important and vitally significant one is respiration and the awareness effortlessly but very effectively be fixed on the breathing process.

## 3. Insomnia Treatment by Quantum Anti-Zeno Effect

The insomnia patient's main trouble is the inability to withdraw the mind from the problematic thoughts and this can effectively be dealt with by introducing another thought on which the attention can be easily fixed. This new thought will be the auxiliary thought-form with a greater awareness and willful attention than the problematic thought-form and it will decrease the frequency with which the problematic thoughts were engaging the attention. The characteristic restlessness of the mind implies that it gets bored of monotony and one-pointedness and if continuously and consciously it is fed with something monotonous it will easily lapse into sleep. Consciously trying to focus on this auxiliary thought-form will gradually bring in the required transition to the fourth state from which the patient will move into the dream and from then on to deep sleep. This is the program of treatment by the ant-Zeno effect. In many situations in physics, it has been observed that a normal transition is inhibited by the QZE and is accelerated by the QAZE in the presence of the auxiliary state. Further as required by the theory of QAZE, the auxiliary state is of higher energy (i.e. frequency) compared to the problem state and admits of transitions back and forth with it. This is precisely the case here but only thing is that we are considering the application in a psychological setting. The four states of consciousness are depicted in Fig. 1a below and in Fig. 1b we show all the transitions as per the QAZE.

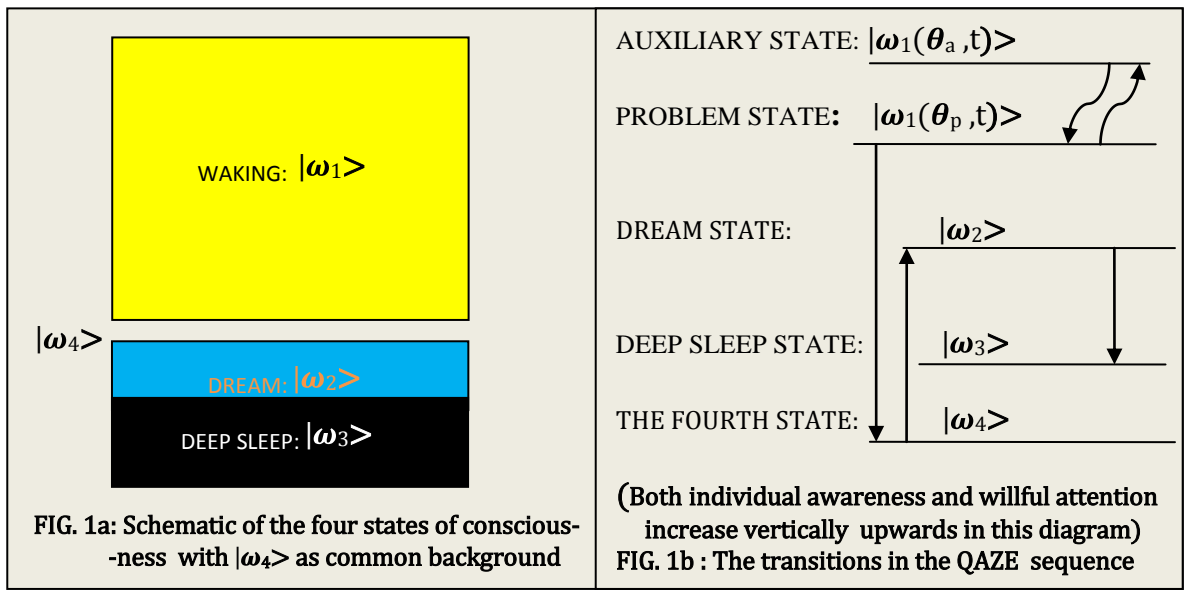

FIG. 1a: Schematic of the four states of conscious--ness with $|\omega_4\rangle$ as common background

(Both individual awareness and willful attention increase vertically upwards in this diagram)
FIG. 1b : The transitions in the QAZE sequence

As shown in Fig.1a, for normal persons, the transition from waking to dream through the thin layer of the fourth state is naturally accomplished without any effort and then deep sleep ensues as a matter of course, while for the insomniac this transition is inhibited by the QZE-type continuous dwelling in the problematic waking state $|\omega_1(\theta_p,t)\rangle$. In Fig. 1b, $|\omega_1(\theta_a,t)\rangle$ is the auxiliary waking state with a different thought-form $\theta_a$, introduced to divert the attention from $\theta_p$ and thus disrupt the QZE. Again, there will be a seeming tussle between $\theta_a$ and $\theta_p$ to engage the patient's continued attention and in the beginning the auxiliary state may seem to be an additional complication! It is almost an implementation of the old adage: "*set a thief to catch a thief!*" But, it works, thanks to the mysterious nature of Quantum phenomena which baffle classical thinking, whether applied to the physical or the psychological phenomena, and one finds to one's surprise that in a matter of about ten minutes to half an hour or so, the patient will be fast asleep via the QAZE transitions!

Now comes the question of the exact nature of the auxiliary thought $\boldsymbol{\theta}_a$, which is of central importance in the entire process. Although $\boldsymbol{\theta}_a$ could be any pleasant thought-form on which the patient likes to dwell upon for a longer period, it turns out that the most effective, unbiased and natural choice is the '*attention on the breath*'. Just gently trying to focus on the normal, natural, relaxed breathing will do the trick! Initially, if need be, one may count the exhalations one by one serially if the disturbance from the thought(s) $\boldsymbol{\theta}_p$ is a bit uncontrollable, but after the first five minutes or so, one may stop counting and try to concentrate only on the natural breathing pattern without bothering much about the frequency of $\boldsymbol{\theta}_p$. In the first few attempts the patient may feel a little tired or bored of the entire exercise, and may, if need be, allowed to have an accompanying mental utterance of some self-chosen holy name or formula (*mantra*) depending on the faith and temperament. This has a very helpful effect but it is not absolutely necessary in all cases. Only a little patience on the part of the patient in participating in the whole therapy will work wonders even if $\boldsymbol{\theta}_a$ is only the 'concentration on the breathing'.

It is to be emphasized that the conscious focusing of attention on $\boldsymbol{\theta}_a$ must not be considered as a fight or a battle with the persistence of $\boldsymbol{\theta}_p$, rather, one should deal with the issue in a very mild and gentle manner allowing the mind its own freedom of dwelling on $\boldsymbol{\theta}_p$ every now and then. The only job is to focus on $\boldsymbol{\theta}_a$, and not to fight a battle with $\boldsymbol{\theta}_p$. As many times as it wonders off to $\boldsymbol{\theta}_p$, so many times very calmly and gently it should be brought back to $\boldsymbol{\theta}_a$ without generating any tension or sense of struggle.

## 4. Practical steps of Anti Zeno-therapy

The most important part of this delicate method of treatment is the preparation stage for the application of the Anti-Zeno therapy to insomnia cases. It is to be emphasized that the three conditions necessary for an early onset of sleep are: (a) a relaxed body (b) a natural rhythmical breathing and (c) a calm, tension-free mind. In one phrase it is 'a total relaxation' of the individual- relaxed body, relaxed breathing and relaxed mind. Therefore, apart from ensuring that the patient has the right kind of cooperative attitude to receive the therapy and a positive attitude towards the efficacy of the therapy the following steps may be noted by the therapist:

- Removal/avoidance of all environmental and physiological factors which have the potential to disturb or delay the onset of sleep i.e. stimulating food, drinks, music and video that excite the mind too much. All sensory inputs throughout the day must invariably be of a soothing nature which helps bring about a calm interior. All kinds of worries, cares and anxieties are to be avoided at all costs.
- Making available all the helpful factors for inducing sleep such as: (a) sleeping on back or on the left side (b) remaining fully engaged in some self-chosen work which the patient enjoys doing in the most relaxed manner possible during the day (c) regularity of the daily schedule of life etc.
- Instructions may be given directly or through a record-player in a commanding voice to start with, followed by a gradual switch over to a very soothing and sweet voice as the patient is guided into deep sleep. The commands in the beginning lessen the vehemence of the problematic thoughts by much, more so if the treatment is in a group. They also have a very

- positive impact on the overall receptivity of the patient and the fixing of attention on the breath becomes much easier.
- In case of individual patients, constant watch may be kept on the patient's eyelids for the onset of dream through REM-sleep, in which case the instructions may be switched off to allow for sleep to ensue.
- In group-therapy sessions, the instructor may switch over to the soothing mode after the first five minutes or so.
- The first few instructions should be on lying flat on the back followed by the step-by-step relaxation of the whole body starting with the toes and ending at the crown of the head.
- The whole session should be in a very friendly and warm environment where the patient feels completely relieved and relaxed and fully at home.

These are some of the very helpful practical hints which are essential for the therapy. With experience the therapist will gain strength and the success rate will be very high. After a week of guided therapeutic sessions the patients should be watched in one or two do-it-yourself sessions before their release. Once the patient gets confidence in the Anti-Zeno practice, insomnia is effectively fully cured.

## 5. Discussion & Conclusion

We have proposed for the first time a novel method of treatment of insomnia exploiting the parallelism of the situation involving the states of consciousness with the physics of Quantum Zeno and the Anti-Zeno Effects, which have been experimentally observed in laboratories in many physical systems. A theoretical objection may be raised regarding the efficacy of the procedure in the event of the continuous observation of the breathing leading to another Quantum Zeno Effect involving this new auxiliary state, in which case the patient effectively has now $\boldsymbol{\theta}_a$ – insomnia in place of $\boldsymbol{\theta}_p$ – insomnia! This apprehension, however, is without basis since breathing is natural in sleep and is therefore not an obstacle to sleep, while the problematic thoughts are an obstacle. As regards the awareness of breaths which is not natural to sleep, since sleep means non-awareness of everything including breathing, the state of the patient then becomes one of "Yogic sleep" or *Yoga-nidra*, [12], which grants all the benefits of sleep (restfulness, rejuvenation, freshness, renewed vigor and vitality etc.) and at the same time takes away all the evils of insomnia (uneasiness, weirdness, heaviness, headache, drowsiness, sloth, stupor, fatigue, etc.), in which case also the patient is cured! Also, as far as the cure is concerned it matters little whether the transition is from $\boldsymbol{\theta}_a$ or $\boldsymbol{\theta}_p$ and whether the transition is direct or through intermediate states.

In fact, the Anti-Zeno Therapy proposed here has been somewhat in practice in the theory and practice of Yoga and Meditation[12]. In particular, the step-by-step whole-body-relaxation is an essential part of the practice of a Yogic posture known as *savasana* (the corpse pose) which very often lands the practitioner in deep sleep. Similarly, 'concentration on the breath' is a preliminary technique in the practice of concentration and meditation with a view to making the mind one-pointed and finally thoughtless. But in both the cases, the lapse to sleep is the most undesirable event and is considered as an obstacle, since the continuity of awareness is of paramount importance in all yogic practices, and the onset of sleep obviously deprives one of that. But, the failure (lapse-to-sleep) of the Yoga-meditation practitioner is verily the 'glorious triumph' for the insomniac! And, it seems that finally there must be some truth in alternative therapies like 'Yoga'.

It is worth noting that the problematic thought-form $\theta_p$ is not a singular form but is usually dressed with two or three other associated thought-forms which succeed in keeping the mind revolving, though centered on $\theta_p$. Thus, there is a difference between $\theta_p$ and the auxiliary thought-form $\theta_a$ which is more of a singular nature and thus easily succeeds in getting the mind bored of monotony thereby facilitating the transition to sleep.

The application of quantum theory to understand the dynamics of the mind has, of course, its own limitations. For one thing, the physical systems investigated in relation to QZE and QAZE are more or less of known eigenstates with known energy eigenvalues, while here the corresponding property is the willful attention (or awareness) which evolves with time and thus the states corresponding to $\theta_p$ and $\theta_a$ do not remain fixed with time. The actual dynamics demands that they gradually shift downwards (see Fig. 1b). It may even so happen that as the attention stabilizes more and more on $\theta_a$ and, consequently, as the mind starts withdrawing itself, $|\omega_1(\theta_a,t)\rangle$ may slide below $|\omega_1(\theta_p,t)\rangle$ and this may bring in another real but beneficial QZE with $|\omega_1(\theta_a,t)\rangle$. In this case the QAZE transition may very well involve the dream state $|\omega_2\rangle$ itself because of its closeness to the auxiliary state. In such a situation, whether a similar Anti-zeno therapy will be the helpful in the cases of dreamful insomnia, where $|\omega_2\rangle$ rather than $|\omega_1\rangle$ is the problematic state, needs further investigation.

## Acknowledgements

The author gratefully acknowledges the impetus, invitation and inspiration from Reem Ali for writing this article.